\documentclass[a4paper,twocolumn,11pt,unpublished]{quantumarticle}
\pdfoutput=1
\usepackage[utf8]{inputenc}
\usepackage[english]{babel}
\usepackage[T1]{fontenc}
\usepackage{amsmath}
\usepackage{amsfonts}
\usepackage{amssymb}
\usepackage{hyperref}
\usepackage{braket}
\usepackage{lipsum}
\usepackage{enumitem}
\usepackage{verbatim}
\usepackage[numbers,sort&compress]{natbib}

\newtheorem{theorem}{Theorem}
\newtheorem{definition}{Definition}
\newtheorem{lemma}{Lemma}
\newtheorem{corollary}{Corollary}
\newtheorem{observation}{Observation}

\begin{document}

\title{Quantum jumbled pattern matching}

\author{Julio Juárez-Xochitemol}
\affiliation{Centro de Investigación Científica y de Educación Superior de Ensenada, Tijuana-Ensenada 3918, 22860 Ensenada, Mexico}
\orcid{0000-0002-5348-0402}
\author{Edgar Chávez}
\orcid{0000-0002-0148-695X}
\affiliation{Centro de Investigación Científica y de Educación Superior de Ensenada, Tijuana-Ensenada 3918, 22860 Ensenada, Mexico}
\maketitle

\begin{abstract}
Let $S_1, S_2 \in \Sigma^*$ strings, we say that $S_1$ {\em jumble match} $S_2$ if they are permutations of each other. Given a text $T$ of size $N$ and a string $S \in \Sigma^*$, the problem of \emph{Jumbled Pattern Matching} (JPM) is to determine all substrings in $T$ jumbled matching $S$. In classical computing, a widespread conjecture is that JPM requires $\Omega(N^{2-\epsilon})$ preprocessing time and space for $O(1)$ query time, or $\Omega(N^{1-\delta})$ query time in the online version, with $\epsilon, \delta >0$. In this paper, we present a quantum algorithm for the online JPM in $O(\sqrt{N})$ time.
\end{abstract}

\section{Motivation}
We can quantify the similarity between two strings employing a variety of metrics. One example frequently used is the edit, or \emph{Levenshtein distance}, which counts the number of insertions, deletions, and substitutions required to convert one string into another; this metric, in particular, can compare strings with different lengths. Another example is the Hamming distance, which measures the number of positions where two same-length strings differ. One less known similarity between same-length strings is the \emph{jumbled} distance. Rather than a distance, it is a discrete semi-metric that verifies if two strings match modulo permutations, or in other words, if their characters and frequencies are the same. The output of the comparison of two strings under this semi-metric will be $0$ for a \emph{jumbled match}, otherwise will be $1$ or not a \emph{jumbled match}.
\\

Jumbled pattern matching (JPM) arises in problems of biocomputing related to pattern analysis, specially in mass spectrometry for data interpretation \cite{Bcker2004SequencingFC}, SNP discovery, composition alignment \cite{Benson2003CompositionA} and bio-sequencing applications in general \cite{Eres2004PermutationPD}.  It is not required to obtain exact matches; instead, we require finding a permuted version of query substrings. A particular example of its application is finding gene orthologs in sequence homology, where orthologue sequences are similar segments of DNA or RNA that come from a branching point called \emph{speciation event}. We can define orthologs as follows:
\paragraph{Orthologs} Given two biological sequences $S$ and $S'$ from two different species, and $d(x,y)$ some appropriate distance measure. If $d(S,S')<k$, with $k$ a threshold, the sequences in comparison identify each other as their closest partners or orthologs. The biological sequences can be a group of genes, amino-acid sequences, nucleotides, among others.
\\

The above definition is indifferent to the position of the genes in the genome, which means that a sequence $G_{1}G_{2}G_{3}$ align with its orthologs $G'_{3}G'_{1}G'_{2}$; however, such an alignment would be undetected in sub-sequence local alignments, where just locally conserved gene clusters are detected \cite{Eres2004PermutationPD}. So, the \emph{jumbled matching} is the proper metric for orthologs detection.

\section{Problem statement}
Let $\Sigma=\{a_{1},..,a_{|\Sigma|}\}$ be an ordered finite alphabet, and a string $s\in \Sigma^{*}$. A Parikh vector counts the frequency of elements that compose a string, i.e. the Parikh vector of some string $s$ has the form $p(s)=\{p_{1},p_{2},...,p_{|\Sigma|}\}$, where $p_{i}=|\{j|s_{j}=a_{i}\}|$ is the number of ocurrences of the $i$-th element of $\Sigma$ . Given  a string $T$ of size $N$ and a substring $q$ of size $m$, both constructed from the same alphabet $\Sigma$, and $m\leq N$. We say that an occurrence of the Parikh vector $p(q)$ in the string $T$ takes place when a substring in $T$ has the same multiplicity of characters of $q$; that is they have the same Parikh vector. 
\paragraph{Jumbled pattern matching.}
Let $\Sigma$ be an alphabet, $T$ a string of size $N$ and $q$ a substring. Given a query Parikh vector $p_{q}$, find all substrings $s\in T$ such that $p_{q}=p_{s}$. 

\section{Classical solutions}
For real applications, we consider the length of the substring $q$, a constant $|q|<<|T|$. We can trivially solve the JPM for a single pattern in $O(N)$ time with a sliding window algorithm, searching exhaustively along $T$ for the substring with the same Parikh vector.
If we want to speed up the query time of JPM, we can preprocess the larger string to build an index; we call this approach \emph{jumbled indexing} (JI). JI seeks the efficient construction of an index with all the substrings of $T$ and a query $q$. In the indexed search, for instance, using the Hamming or Levenshtein metrics, the goal is to find those strings with minimal distance to the query; meanwhile, in JI, the goal will be to find all the strings that jumbled match the query. We build the index before queries; after that, the searching needs constant time. 
\\

Consider a single text $T$, formed by all size $m$ substrings, where $m$ is the size of the query. There are two trivial solutions:
\begin{itemize}
\item \emph{Without an index}, in linear time with a sliding window algorithm.
\item \emph{With an index}, using quadratic time and space to construct the index of all substrings; i.e. build dictionaries for sizes $2,3,4,...,N$, each dictionary will represent a different \emph{Parikh vector}. The query time will be constant.
\end{itemize} 
The JPM problem is suspected to have quadratic complexity using an index, and linear complexity without an index. Although authors in \cite{Amir2014OnHO} claimed that it was not possible to construct an index for JPM in less than $O(n^2)$ time, the proof was based on \emph{3-SUM} hardness assumption, in \cite{Jrgensen2014ThreesomesDA} the \emph{3-SUM} hardness assumption was refuted. Thus, the reduction in \cite{Amir2014OnHO} need to be updated for JPM and several other string matching problems. 
\\

Below we present the Quantum Pattern Matching Algorithm in \cite{Mateus2005QuantumPM}, essencial part of our quantum approach to JPM problem, its complexity in a subsequent section and, finally, how we use the initial entangled state for getting a superposition of Parikh vectors by two different ways. In such a way that a Grover's iteration can be applied to get a permuted sub-pattern position almost certainly. 

\section{Quantum Pattern Matching}

In \cite{Hariharan2003StringMI} and \cite{Mateus2005QuantumPM} authors show quantum algorithms for the pattern matching problem from distinct perspectives. In \cite{Hariharan2003StringMI} authors use particular query functions dependent on the searched pattern, whereas, in \cite{Mateus2005QuantumPM} query functions are independent of the pattern. Since query functions dependent on the searched pattern would result in different settings of a quantum circuit for different patterns, this makes it difficult to implement in realistic situations. From a complexity point of view, query functions can be prepared beforehand matching the number of oracles with the size of the alphabet. In real applications of pattern matching this can be a reasonable assumption; for instance, looking for substrings in DNA patterns, where the alphabet size is constant. In that way, we shortly present the quantum pattern matching algorithm from P. Mateus \cite{Mateus2005QuantumPM}; focusing on the preparation, the critical component of our proposed quantum algorithm.
\\

The central part of Mateus algorithm uses Grover's routine for multiple elements encoded in states of a  Hilbert space $\mathcal{H}$ \cite{Boyer1996TightBO}. We can summarize the routine as follows: Let $N$ the size of a large string (database) encoded in $\mathcal{H}$, we apply a query operator over the element we are looking for, marking it by changing its phase; we amplify the amplitude of the marked state in a subsequent step. We repeat this process $O(\sqrt{N})$ times, to finally detect the state with non negligible probability. This algorithm allows searching for any number of different patterns in a given string, the algorithm returns the position of the closest sub-pattern to a given query pattern of size $M$ with non-negligible probability in $O(\sqrt{N})$ time, where $N$ is the size of the large string. Below we state some definitions and the problem solved in \cite{Mateus2005QuantumPM} and \cite{Hariharan2003StringMI}.
\\
\begin{definition}{\bf Hamming Distance}
Let be $u,v \in \Sigma$ two strings of the same size, the Hamming distance between $u$ and $v$ is defined as the number of positions where they differ.
\end{definition}
\begin{definition}{\bf Pattern Matching with at most $k$ differences}
Given $w,p \in \Sigma^{*}$ strings of size $|w|=N$ and $|p|=M$, with $N>>M$, take an additional parameter $k$. Get all indices $i$, for $1\leq i\leq (N-M+1)$ for which the hamming distance between  $w_{i}w_{i+1},...,w_{i+m-1}$ and $p_{1}p_{2},...,p_{m}$ is less or equal to $k$.
\end{definition}
\begin{definition}{\bf Closest pattern matching}
Given the pattern matching problem with at most $k$ differences, return the pattern with lower $k$.
\label{definition:CPM}
\end{definition}

\subsection{Preparation}
The system's initial entangled state will capture an analogy from the classical sliding window algorithm, in which we scan the large string with a window of size $M$. We want a state where the second symbol of $q$ occurs just after the first, the third after the second, and so on. The proper preparation satisfying the above requirement is as follows:

  \begin{equation}
  \label{eq5}
\begin{split}
\ket{\psi_{0}}= \\
\frac{1}{\sqrt{N-M+1}}\sum_{k=1}^{N-M+1}\ket{k,k+1,...,k+M-1}
\end{split}
  \end{equation}

Notice that we start with the state space $\mathcal{H}^{\otimes M}$, and just after the preparation, the search space shrinks to a subspace of dimension $N-M+1$, where only the $N-M+1$ states survive.
\\

To correctly apply the Grover iterator in a subsequent step over the survive states, we define the proper query function:
 \begin{equation}
  \label{eq6}
f_{\sigma}(i) = \left\{ \begin{array}{lcc}
             1 & \text{if the }i-\text{th letter of }w \text{ is }\sigma\\
             0 & \text{otherwise}
             \end{array}
   \right.
  \end{equation}
In this approach, we need particular query operators, one for each element of the alphabet; so, the shifting phase will be performed by the next unitary transformation:
\begin{equation}
  \label{eq7}
U_{\sigma}\ket{k}=(-1)^{f_{\sigma}(k)}\ket{k}
  \end{equation}
while the amplitude amplification operator is the usual one from Grover's routine.
\\

Every symbol of the pattern, selected at random, and its correspondent query operator are applied to the proper position. Each match will produce an amplification, with each match resulting in a higher probability of detection at the measurement stage. If $N>>M$, as in real applications, sampling about $\sqrt{N}$ times the $M$ elements of the pattern will produce an amplification significant enough to be measured with high probability. Finally, we obtain the position of the closest match to $p$ by performing a measurement of the system over the basis $B=\{\ket{1}, \ket{2},...,\ket{N}\}$. The number of iterations needed to observe a match with non-negligible probability is, as expected, $O(\sqrt{N})$.
\\

Regarding to the amplification stage, this is performed by applying the usual Grover diffusion operator $D=D_{N}\otimes I^{M-1}$ to the total state, where:
\begin{equation}
    D_{N}=2(\ket{\psi}\bra{\psi})-I
\end{equation}
and $\ket{\psi}$ is a uniform superposition obtained with a common Hadamard operator, it is given by
\begin{equation}
    \ket{\psi}=\Sigma_{i=1}^{N}\frac{1}{\sqrt{N}}\ket{i}
\end{equation}
\\

The next pseudo-code summarize the quantum pattern matching algorithm:
\\
\\
Input: $w$, $p \in \Sigma^{*}$
\\
Output: $m \in [1,N-M+1]$
\\
Quantum variables: $\ket{\psi}$
\\
Classical variables: $r,i,j \in \mathbb{N}$
\begin{enumerate}
\item Choose $r\in [1,\lfloor \sqrt{N-M+1}\rfloor]$ uniformly,
\item Set $\ket{\psi_{0}}=\frac{1}{\sqrt{N-M+1}}\sum_{k=1}^{N-M+1}\ket{k,k+1,...,k+M-1}$;
\item for $i=1$ to $r$:
\begin{enumerate}
\item Choose $j\in [1,M] $ uniformly.
\item Set $\ket{\psi}=I^{\otimes j-1}\otimes Q_{p_{j}}^{w}\otimes I^{\otimes M-j}\ket{\psi}$ (phase shifting);
\item Set $\ket{\psi}=(D\otimes I^{\otimes M-1})\ket{\psi}$ (Grover's diffusion).
\end{enumerate}
\item m $\Leftarrow$ Measure $\psi$ over the base $B$
\end{enumerate}
\section{Complexity}
The circuit complexity of the algorithm for quantum pattern matching is $O(N^{3/2}\log^{2}(N)\log(M))$ without taking into account the initial preparation; meanwhile, for a classical circuit, the complexity it turns out to be $O(MN^{2})$. With regard to the initial preparation, it can be achieved by implementing Pauli-X gates. To get the particular sequence that encodes the order of the symbols of $p$ we must create the entanglement state $\Sigma_{i=0}^{2^{s}-1}\ket{i,i}$, it can be performed by applying  a sequence of Pauli-X gates $O(\log^{2}(N-M))$ multi-controlled Pauli-X gates; in the same way, this gates can be obtained using $O(\log(N-M))$ C-NOT and Pauli-X gates. Thus, the circuit complexity of the initial entanglement state is $O(\log^{3}(N-M))$.
\\
\begin{figure}[t]
  \centering
  \includegraphics[scale=0.36]{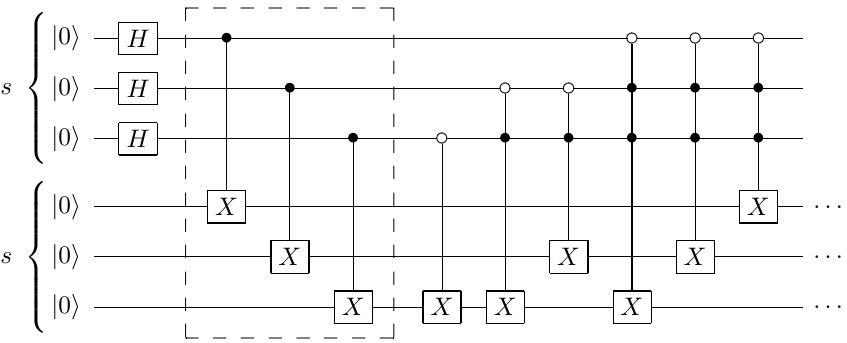}
  \caption{Circuit that generates the initial preparation given by \eqref{eq5}. In a first step we need to create a superposition of the $s$ qubits, Hadamard gates perform the operation. The box groups Pauli-X gates for the entanglement state $\Sigma_{i=0}^{2^{s}-1}\ket{i,i}$; then we apply a sequence of $O(\log^{3}(N-M))$ multi-controlled Pauli-X gates. The first and second set of $s$ lines represents the procedure for the first and second symbol of the pattern; so, we must iterate another $M-3$ times, giving us an overall complexity of $O(M\log^{3}(N-M))$.}
  \label{fig:InitialSuperpositionCore}
\end{figure}
\\

It results that the algorithm has an efficient compile time of $O(N\log^{2}(N))\times |\Sigma|$ and a total run time of $O(M\log^{3}(N)+N^{3/2}\log^{2}(N)\log(M))$, where the execution time is the compile stage plus the Grover's iteration. As the algorithm can be used to make multiple queries, it can be thought as based on a \emph{compile once, run many times} approach. So, the query complexity of the quantum pattern matching algorithm is $O(\sqrt{N})$ considering the size $M$ of the pattern much smaller than the size $N$ of the database; this is, in the realistic and relevant application of the algorithm.
\\
\\
A complete and detailed implementation of the algorithm can be found in the original article by P. Mateus and Y. Omar \cite{Mateus2005QuantumPM}.
\section{Quantum Jumbled Pattern Matching}
\subsection{First approach}
In \cite{Mateus2005QuantumPM} the authors propose an algorithm to solve the Closest Pattern Matching, as in Definition \ref{definition:CPM}. We will use the same preparation of the above algorithm. The input of our system is the pattern and the text string, as the output, we will detect the position of a jumbled match almost certainly.
\\

We will encode the string in a Hilbert space $\mathcal{H}$, with basis $B=\{\ket{1},\ket{2},...,\ket{N}\}$. Before we can apply the Grover iterator we need to transform the input into a form such that each substring in superposition \eqref{eq5} is an element from a new alphabet $\Sigma'$, where $\Sigma'$ is the alphabet of the possible Parikh vectors. Clearly, two substrings in superposition that jumbled matches will be converted into the same Parikh vector; so, the new alphabet will have a size $|\Sigma'|\leq N-M+1$.
\\

Each state in the superposition (\ref{eq5}) that represents the substring in the sliding window will be translated to a Parikh vector by an operator $U_{p}$  as follows:
\begin{equation}
    U_{p}\ket{k,k+1,...,k+M-1}= \ket{P_{k}}
\end{equation}
where $P_{k}$ is the correspondent Parikh vector.
\\
\begin{figure}[t]
  \centering
  \includegraphics[scale=0.29]{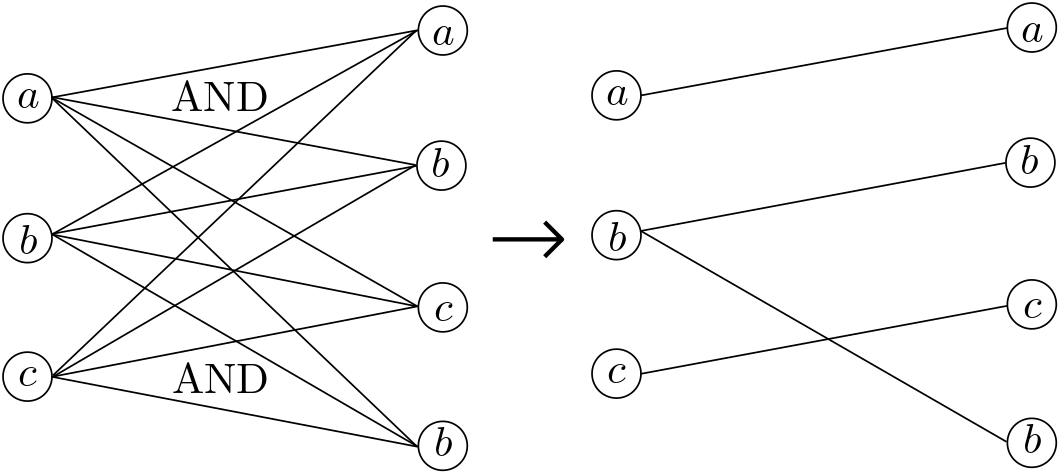}
  \caption{Example of a transformation operator as a bipartite graph. The substring in superposition is on the left, and the original alphabet on the right; each edge represents \emph{AND} operations.}
  \label{fig:transformationOperator}
\end{figure}

\begin{figure}[t]
  \centering
  \includegraphics[scale=0.315]{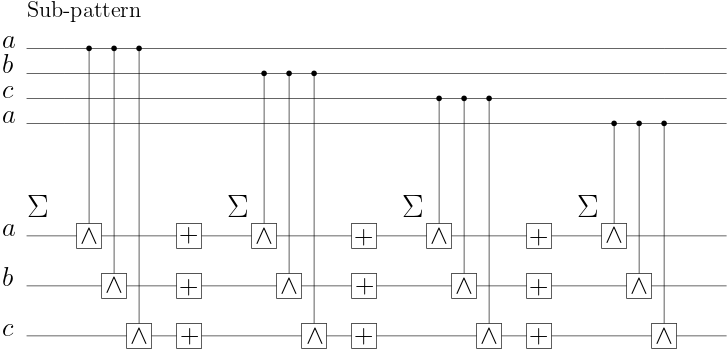}
  \caption{High-level construction of the translate operator presented in Figure \ref{fig:transformationOperator}, the gates $\land$ are reversible $AND$'s and the $+$ are reversible sum gates.}
  \label{fig:Operator_2}
\end{figure}

The first piece of the transformation operator is a complete bipartite graph $U_{p}=(\Sigma,S)$ with $S$ a placeholder for substrings of size $|p|$ (see Figure \ref{fig:transformationOperator}). The transformation operator depends on the query pattern $p$; as the pattern is given in the problem, the corresponding operator is prepared along with the text. In Figure \ref{fig:transformationOperator} we can see an example of this operator as a bipartite graph, where the substring in superposition is on the right and the original alphabet on the left; each edge represents a logical AND, after applying the AND operation the survivor edges count the multiplicities of the elements of $\Sigma$ in the substring. A high-level construction of this translation operator can be visualized in Figure \ref{fig:Operator_2}.
\\

Before applying the Grover's iterator we present a second approach for the translation from the original alphabet to the Parikh vectors' alphabet.
\subsection{Second approach}
In this second approach we start from a different encoding for the $T$ string and the $q$ pattern, so that every element of the original alphabet $\Sigma$ corresponds to a single prime number. The encoding is as follows:
\\

Given an ordered set of prime numbers $\{p_{1},p_{2},...,p_{|\Sigma|}\}$, every element $\sigma\in \Sigma$ is rewritten as
\begin{equation}
   \begin{matrix}
   \sigma_{1} \Rightarrow p_{1}\\
   \sigma_{2} \Rightarrow p_{2}\\
   \vdots \\
   \sigma_{|\Sigma|} \Rightarrow p_{|\Sigma|}\\
   \end{matrix}
\end{equation}
Thus, this encoding will make $w$ and $p$ strings made up of prime numbers; there will be no extra cost for the quantum algorithm as the encoding is given together with the problem. Before the preparation in \eqref{eq5} everything remains the same.
\\
\paragraph{Parikh vector superposition}

Following the recipe of the initial Mateus preparation \eqref{eq5}, now we have a superposition where each \emph{window} or sub-pattern will be constructed from the elements of $\Sigma$ of prime numbers. The next step is to convert every sub-pattern in its corresponding Parikh vector, note that every sub-pattern now is a sequence of numbers. Consider the following theorem:

\begin{theorem}
Let be $\alpha\in \Sigma^{*}$ and $\beta\in \Sigma^{*}$, with $\Sigma=\{2,3,5,...,P_{|\Sigma|}\}$ an alphabet formed by the set of the first $|\Sigma|$ prime numbers.
It holds that the Parikh vectors of both strings are equal if, and only if the product of the characters that make up the strings are the same, i.e:
\begin{equation}
P(\alpha)=P(\beta) \iff \prod_{i} \alpha_{i}=\prod_{i} \beta_{i}
\end{equation}
\end{theorem}

Based on this theorem, We define an operator that takes a sub-pattern of the superposition in \eqref{eq5} as input and calculates its product, product that turns out to be its corresponding Parikh vector. 
\begin{definition}
Let  $Q_{\Pi}$ be the product operator that applied on the sub-pattern $\ket{\alpha_{k},\alpha_{k+1},...,\alpha_{k+M-1}}$ generates the product of its characters:
\begin{align*}
    Q_{\Pi}\ket{\alpha_{k},\alpha_{k+1},...,\alpha_{k+M-1}}\\ 
    =\ket{\alpha_{k}\cdot (\alpha_{k+1})\cdots (\alpha_{k+M-1})}
\end{align*}
\end{definition}

We can apply this operator in a single step over the Mateus superposition, getting a superposition of Parikh vectors as the output:
 \begin{equation}
  \label{eq9}
\ket{\psi_{1}}= 
\frac{1}{\sqrt{N-M+1}}\sum_{k=1}^{N-M+1}\ket{P_{k}}
  \end{equation}
possibly, with some $j,l$ such that $P_{s}=P_{l}$. The  new alphabet  is explicitly stated as $\Sigma'=\{0,1,\dots,|p|\}$.
\\

We can use the Grover iterator for multiple occurrences as in \cite{Boyer1996TightBO}. Unlike the algorithm for closest pattern matching, where the Grover iterator could amplify the amplitude of substrings with partial ocurrences, the last translation allow us to increase the probabilities of the elements in superposition with the correct Parikh vector. To correctly apply the Grover iterator we slightly modify the query function as follows:
\begin{equation}
  \label{eq10}
f_{p}(i) = \left\{ \begin{array}{lcc}
             1 & \text{if the }i-\text{Parikh vector of T} \text{ is }p\\
             0 & \text{otherwise}
             \end{array}
   \right.
  \end{equation}

so that, in the redefined alphabet $\Sigma'$ could exist $j,k$, such that $f_{p}(j)=f_{p}(k)$.
\\
\section{Complexity}
The complexity of the algorithm will be identical to the pattern matching algorithm in \cite{Mateus2005QuantumPM}. The only addition to the analysis is the call to the operator that generates the product of the elements of the sub-pattern (or the obtaining of the Parikh vector in the first approach), in which, using the quantum parallelism operates in a single call, simultaneously over every sub-pattern in superposition. After the translation to Parikh vectors the application of the Grover's iterator is performed $O(\sqrt{N})$ times; therefore, the query complexity of our algorithm is $O(\sqrt{N})$. As in  \cite{Mateus2005QuantumPM} we use the approach of  \emph{compile once, run multiple times}, where the initial preparation is done just one time, offering the possibility of carrying out many different searches, we justify the query complexity of our algorithm under the same criteria than in Mateus and Omar algorithm, removing the subtleties of the machine (circuit) that would implement this algorithm.

\end{document}